\documentstyle[aps,prd,multicol]{revtex}

\begin{document}
\title{The cosmological constant, neutrino masses and parity symmetry in 6+1 dimensions}
\author{ Jos\'e  Antonio Martins Sim\~oes \thanks{E-mail: simoes@if.ufrj.br}\\
Instituto de F\'\i sica\\
Universidade Federal do Rio de Janeiro, RJ, Brazil \\}
\maketitle
\begin{abstract}
\par
In this paper we propose that the cosmological constant scale and neutrino masses have a common origin: a new spontaneously broken scalar field. This hypothesis is implemented in a six-space dimensional model that restores the parity symmetry. This new 6-dimensional parity symmetry gives a natural mechanism for the smallness of the vacuum energy density.
\end{abstract}
\vskip 0,5cm
 
\begin{multicols}{2}

It is well known that we have two very small scales in present day physics. One is the cosmological constant scale and the other is the recent experimentally observed neutrino mass. 
\par
It is an old problem  \cite{WEI} the fact that the cosmological constant term in Einstein's equation is related to the vacuum energy density with a small value $\rho_{vac}=(\lambda)^4$ with $\lambda=10^{-3}eV$ . The problem is that all the known physical scales give an enormously larger number. It is generally expected that some symmetry will help in understanding why all the known physical mass scales such as Fermi's scale, QCD, GUT, supersymmetry and so on, do not contribute to $\rho_{vac}$. Yet, a detailed analysis done by Weinberg \cite{WEI} more than a decade ago found no simple explanation for this property. A more recent analysis shows that this situation has not changed \cite{WE2}.
\par
On the other hand, the recent experimental results on solar and atmospheric neutrino properties \cite{NIS} confirm that neutrinos have small masses and oscillate. There is some theoretical dependence on the interpretation of these results. Oscillation experiments give only $\Delta m^2$. But it is consistent with the experimental data \cite{CON} that neutrino masses are on a scale $ m_{neutrino}=10^{-2}-10^{-4}$ eV. 
\par
The fact that these two scales are very close, if not identical, has so far  been considered as a simple coincidence.
Our first hypothesis is that the cosmological constant scale and neutrino masses have a common origin: they result from a new scalar field with  a  spontaneously broken scale  $\lambda=10^{-3}eV$ . It is very simple to generate Majorana and/or Dirac masses from singlet scalar fields, but this hypothesis alone is not sufficient to take into account of the cosmological constant problem.
\par
 We now turn our attention to the idea that  nature can have more than three spacial dimensions.  There are theoretical motivations for  enlarging the number of dimensions in order to incorporate gravity with the other gauge forces in a consistent way. The possibility of phenomenological tests of these ideas at the TeV scale is a promising point. If extra dimensions are non-compact it is possible to localize our known fundamental fields on a thin three-dimensional brane. The possibility of new space dimensions opens an interesting mechanism for the smallness of the vacuum energy density. The usual three dimensional integration over the vacuum energy density must be enlarged for the new extra dimensions. The vanishing of some of the known physical scales could then be a consequence of the "n" dimensional space properties and not only of internal symmetries.
\par
In this paper we will put forward the possibility that  new extra 3-space dimensions $ {\bf y_i}(i=4,5,6) $ could be connected to our known 3-dimensional $ {\bf x_j} (j=1,2,3)$ world by a discrete  six-dimensional parity transformation  ${\bf x_i} \leftrightarrow  {\bf -y_j}$. That parity symmetry is connected with the number of space dimensions can be seen in a simple example. In our three space dimensional world, spatial reflections and space rotations are different, unconnected  transformations. Also, for some reason not explained in the standard model scenario, parity is not a symmetry of nature. But if we turn off one space dimension, and restrict physics in a two-dimensional world, then space reflections and rotations are identical operations and we have no place for parity violation. If we now follow the inverse road, and enlarge the number of space dimensions, we can expect that in higher space dimensions parity is to be conserved. It is the confinement in a three dimensional brane that makes parity to break. 
\par
In this scenario, ordinary matter must be trapped in our three dimensional $\bf x_i$ brane. The left-right 6-dimensional symmetry is restored by attaching new fermions with opposite chirality in a new $\bf y_j$ brane.
\par
The most appealing feature of this model is that we have a natural mechanism that vanishes the contribution of the known physical scales to the cosmological constant. All the vacuum energy densities must be integrated over the six-dimensional coordinates. By the 6-parity symmetry the new volume is "negative". If we postulate that the fields  that are to be broken at the known physical scales are even under 6-dimensional parity, then their contribution vanishes identically. The only exception to this rule is a new scalar field "${\Phi}$ " that breaks at a scale $\lambda=10^{-3}eV$. It must be a pseudo-scalar and it is the only field that contributes to the cosmological constant. This pseudo-scalar field  is a natural candidate to implement the idea of quintessence \cite{WE2,PEE,IOA}.
\par
In order to generate a solution to the hierarchy problem, some models \cite{RAN} were proposed with an orbifold geometry that locates a positive energy brane at one point and a negative energy brane at the second point. As a consequence we have a positive cosmological constant at one brane and a negative one at the other brane. In our model the "negative" cosmological constant is a natural result from the 6-parity symmetry and we must have exactly $\lambda_{\bf x} \leftrightarrow - \lambda_{\bf y}$, without any fine tuning imposition.
\par
It is known that extra dimensions \cite{GER} have a profound effect on the gauge coupling unification. In ref.8, the authors developed a general approach to estimate the cut-off dependence of the gauge couplings $\alpha^{-1}_{i}(\Lambda)$ due to new extra dimensions. This scheme, very similar to the renormalization group technique, includes quantum corrections to the standard model gauge couplings. If we take the standard model three fermionic generations plus one Higgs doublet at one brane and three new generations with opposite chirality  plus a new Higgs doublet at the other brane we  have $(b_1, b_2, b_3)=(-41/10, -1, 3)$. In the minimal scenario of no chiral fermions Kaluza-Klein excitations we have $(\tilde b_1, \tilde b_2, \tilde b_3)=(1/10,-41/6,-21/2)$. With these values for the $b_i$ coefficients, the extra dimensions at a  scale $\mu_{0}=R^{-1}=1 TeV$ and the approach of ref. 8, we have the results shown in figure 1 for the gauge coupling evolution with a cutoff $\Lambda$.
\par
The model has far reaching consequences. First of all, we can have a physical insight into the nature of extra dimensions.
String theory requires that in order to have anomaly cancelation there must be at least  six or seven new extra dimensions. In order to prevent anomalies we can include a new 3-brane that must not be connected by an n-dimensional parity to the previous $\bf x_i$ and $\bf y_j$ branes. In this picture, we have three 3-space branes, two of then connected by a parity symmetry. Gravitational and  neutrino fields could connect the bulk of this nine-space dimensional world. An important point is that all fields that contribute to the vacuum structure must be defined in both  $\bf x_i$ and $\bf y_j$ branes. This will imply in new restrictions on phenomenological models. As this point probably  has no unique answer, we will not pursue any further along these lines.
\par
Neutrino masses and oscillations \cite{DIE} can be generated in different ways. If we introduce a right-handed singlet in our $\bf x_i$ brane, then the 6-parity symmetry requires a left-handed neutrino singlet in the $\bf y_j$ brane. One can then generate Dirac mass terms, Majorana mass terms and combinations of both that are to be coupled to ${\Phi}$. 
\par 
With a small number of extra dimensions, the model also allows to address  fermion mass hierarchy \cite{ARK}, charge quantization \cite{PIR} and supersymmetry breaking \cite{ANT}. 
\par 
In conclusion, the 6-parity symmetry can give a deeper insight into the nature of extra dimensions. Besides restoring the left-right symmetry it gives a new perspective on the connection between   the cosmological constant scale and neutrino masses.

{\it Acknowledgments:} This work was partially supported by the Brazilian agencies CNPq and FAPERJ.We thanks F.M.L. Almeida Jr. and Y.A. Coutinho for their help during the preparation of this article.

\vspace{1cm}
\LARGE
{\bf Figure Caption}
\normalsize
\begin{enumerate}
\item Cut-off dependence of the Standard Model couplings with 3 extra space dimensions.
\end{enumerate}

\end{multicols}
\end{document}